\documentclass{IEEEtran}
\usepackage{amssymb}
\usepackage{amsfonts}
\usepackage{amsmath}
\usepackage{algpseudocode}
\usepackage{algorithm}

\usepackage{caption}
\usepackage{bm}
\usepackage{graphicx,subfigure,cite,multicol,multirow,diagbox,booktabs,array}
\usepackage{stfloats}

\usepackage[dvips]{color}
\usepackage{float}
\usepackage{subfig}
\ifCLASSINFOpdf
\else
\fi
\begin{document}
\title{Deep Learning Based DOA Estimation for Hybrid Massive MIMO Receive Array with Overlapped Subarrays}
\author{Yifan Li, Baihua Shi, Feng Shu,~\emph{Member},~\emph{IEEE}, Yaoliang Song,~\emph{Senior Member},~\emph{IEEE}, and Jiangzhou Wang,~\emph{Fellow},~\emph{IEEE}
\thanks{This work was supported in part by the National Natural Science Foundation of China (Nos. 61271331, 61271331, 62071234, 62071289, and 61972093), the Hainan Major Projects (ZDKJ2021022), the Scientific Research Fund Project of Hainan University under Grant KYQD(ZR)-21008 and KYQD(ZR)-21007, and the National Key R\&D Program of China under Grant 2018YFB180110 (Corresponding authors: Feng Shu and Yaoliang Song).}
\thanks{Y. Li, B. Shi and Y. Song are with the School of Electronic and Optical Engineering, Nanjing University of Science and Technology, Nanjing 210094, China. (e-mail: liyifan97@foxmail.com).}
\thanks{F. Shu are with the School of Information and Communication Engineering, Hainan University, Haikou 570228, China. (e-mail: shufeng0101@163.com).}
\thanks{J. Wang is with the School of Engineering, University of Kent, Canterbury CT2 7NT, U.K. (e-mail: j.z.wang@kent.ac.uk).}
}\maketitle
\begin{abstract}
To improve the accuracy of direction-of-arrival (DOA) estimation, a deep learning (DL)-based method called CDAE-DNN is proposed for hybrid analog and digital (HAD) massive MIMO receive array with overlapped subarray (OSA) architecture in this paper. In the proposed method, the sample covariance matrix (SCM) is first input to a convolution denoise autoencoder (CDAE) to remove the approximation error, then the output of CDAE is imported to a fully-connected (FC) network to get the estimation result. Based on the simulation results, the proposed CDAE-DNN has great performance advantages over traditional MUSIC algorithm and CNN-based method, especially in the situations with low signal to noise ratio (SNR) and low snapshot numbers. And the OSA architecture has also been shown to significantly improve the estimation accuracy compared to non-overlapped subarray (NOSA) architecture. In addition, the Cramer-Rao lower bound (CRLB) for the HAD-OSA architecture is presented.
\end{abstract}
\begin{IEEEkeywords}
Direction-of-arrival (DOA), massive MIMO array, overlapped subarray, deep learning, Cramer-Rao lower bound (CRLB).
\end{IEEEkeywords}
\section{Introduction}
Direction-of-arrival (DOA) estimation has been an important research direction in the areas of wireless communications, radar, sonar, etc, for a long time. With the development of 5G, the massive MIMO system has been studied extensively. However, the realization of the traditional full digital system requires a high hardware complexity, so the hybrid analog and digital (HAD) system was considered as an alternative\cite{sohrabi2016hybrid}. Then the DOA estimation problem for the HAD system was discussed in \cite{chuang2015high} and \cite{shu2018low}. In addition to the common architecture, various special architecture were considered in \cite{zhang2021direction}\cite{9877909}, and the overlapped subarrays (OSA) architecture in \cite{song2017overlapped} was proved to have better beamforming performance than nonoverlapped subarrays (NOSA) architecture.

Traditional DOA estimation methods are mainly divided into two categories: parameter estimation-based methods and subspace methods\cite{tuncer2009classical}. The first category contains nonlinear least-square (NLS) estimator and maximum likelihood (ML) estimator, while the classical subspace methods include MUSIC, ESPRIT, root-MUSIC, etc. Recently, the deep learning (DL)-based methods have been new choices for solving the DOA estimation problems, since they have lower complexity than parameter estimation-based methods and higher accuracy than subspace methods. In \cite{liu2018direction}, a deep neural network (DNN) was proposed for DOA estimation with array imperfections. The convolution neural network (CNN) was also used in \cite{papageorgiou2021deep} for the improvement of accuracy in the low signal-to-noise-ratio (SNR) regime. And \cite{hu2019low} gave a DNN-based method for the DOA estimation with HAD massive array. The DOA estimation problem with low-resolution ADC was considered in \cite{shi2022impact} and \cite{9760509}. A fast ambiguous elimination method for DOA estimation was also proposed in \cite{chen2021fast}.

The autoencoder (AE) is a kind of neural network, and it is trained to copy the input to the output. In \cite{liu2018direction}, the AE was used to map the inputs into the corresponding DNN network. When the input data contains noise, we can obtain the noiseless data by denoising autoencoder (DAE) proposed by \cite{vincent2008extracting}. And replace the hidden layers in the DAE with the convolution layers, then the convolution denoising autoencoder (CDAE) which is widely used in the field of image processing is obtained\cite{mao2016image}.

In this letter, the DL-based DOA estimation method CDAE-DNN is proposed for the HAD massive MIMO receive array with overlapped subarrays. Our main contributions are summarized as follows:
\begin{enumerate}
\item To improve the accuracy of DOA estimation for hybrid massive MIMO array, the HAD-OSA architecture is implied in this work. As the number of elements in each subarray is the same as the NOSA architecture, the OSA architecture has more RF chains to achieve a larger virtual aperture, and it can get more accurate estimation results. The simulation results also show that the OSA has better performance than NOSA when the SNR and the number of snapshots are low. And the CRLB for the special HAD-OSA architecture is also given in this work.
\item In order to solve the DOA estimation problem for HAD-OSA architecture, a DL-based method called CDAE-DNN is also proposed in this letter. In this method, the input data is first imported to the CDAE for clearing errors, and then the FC network is employed to perform the multi-classification task. Comparing the simulation results of the proposed CDAE-DNN, MUSIC and CNN in \cite{papageorgiou2021deep}, it is obvious that the CDAE-DNN has significant advantages over the other methods. Especially, the CDAE-DNN can achieve the accuracy lower bound at SNR=-10dB when $N=100$, but MUSIC and CNN in \cite{papageorgiou2021deep}  need SNR$\geq$-5dB to achieve the same bound.
\end{enumerate}

\emph{\rm{\textbf{Notation}}:}Matrices, vectors, and scalars are denoted by letters of bold upper case, bold lower case, and lower case, respectively. Signs $(\cdot)^T$ and $(\cdot)^H$ represent transpose and conjugate transpose. $\mathbf{I}$ and $\textbf{0}$ denote the identity matrix and matrix filled with zeros. $\textrm{Re}\{\cdot\}$ and $\textrm{Im}\{\cdot\}$ represent the real part and imaginary part of a complex number.
\begin{figure}[ht]
  \centering
  \includegraphics[width=0.35\textwidth]{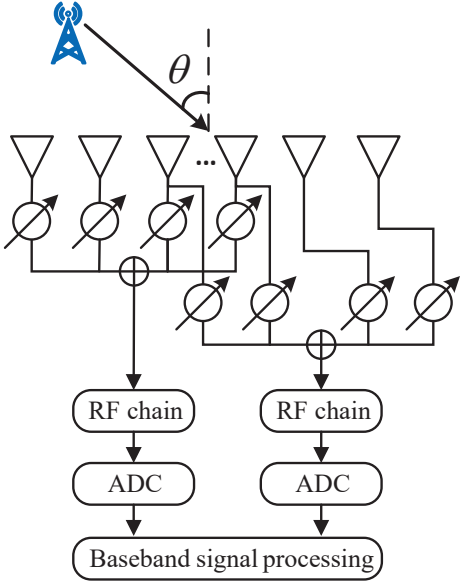}\\
  \caption{System model.\label{system_model}}
\end{figure}
\section{System Model}
Consider $Q$ far-field narrow-band signal received by a massive MIMO receiver equipped with an $M$-element uniform linear array (ULA). The $q$th signal is expressed as $s_q(t)e^{j2\pi f_c t}$, where $s_q(t)$ is baseband signal and $f_c$ is carrier frequency. As shown in Fig.\ref{system_model}, we divide this array into $K$ overlapped subarrays, each of which is connected to a RF chain and composed of $M_s$ antenna elements. The number of overlapped antennas between two adjacent subarrays is denoted by $\Delta M_s$, and we can get $M=KM_s-(K-1)\Delta M_s$. As special cases, when $\Delta M_s=M$ or $\Delta M_s=0$, this array is with fully-connected hybrid architecture or NOSA hybrid architecture.
After down conversion and analog/digital conversion, the received baseband signal is formulated as
\begin{equation}
\begin{aligned}
\mathbf{y}(n)&=[y_1(n),\cdots,y_K(n)]^T\\
&=\mathbf{W}^H\mathbf{A}(\boldsymbol{\theta})\mathbf{s}(n)+\mathbf{W}^H\mathbf{v}(n)\\
\label{received_signal}
\end{aligned}
\end{equation}
where $\mathbf{s}(n)$ is assumed as a stationary zero-mean Gaussian random process. $\mathbf{A}(\boldsymbol{\theta})=[\mathbf{a}(\theta_1),\cdots,\mathbf{a}(\theta_Q)]\in \mathbb{C}^{M\times Q}$ is the array steering matrix, where $\boldsymbol{\theta}=[\theta_1,\cdots,\theta_Q]^T$  and $\mathbf{a}(\theta_q)=[1,e^{j\frac{2\pi}{\lambda} d\sin \theta_q},\cdots,e^{j\frac{2\pi}{\lambda} (M-1)d\sin \theta_q}]^T$. $\mathbf{v}(n)\sim\mathcal{CN}(\boldsymbol{0},\sigma_v^2\mathbf{I}_M)$.
The complete analog beamforming matrix $\mathbf{W}\in \mathbb{C}^{M\times K}$ is expressed as
\begin{equation}
\mathbf{W}=\begin{bmatrix}
w_{1,1}&\quad &\quad&\quad\\
\vdots&w_{2,1}&\quad&\quad\\
w_{1,M_s}&\vdots&\ddots&w_{K,1}\\
\quad&w_{2,M_s}&\quad&\vdots\\
\quad&\quad&\quad&w_{K,M_s}
\end{bmatrix}
\end{equation}
where $M_k=(k-1)(M_s-\Delta M_s)$, $w_{k,m}=\frac{1}{\sqrt{M_s}}e^{j\alpha_{k,m}}$, $\alpha_{k,m}$ is the corresponding phase of $m$th phase shifter in $k$th subarray and it is clear that two adjacent columns of $\mathbf{W}$ are overlapped.

The received signal of $k$th subarray is expressed as
\begin{equation}
\begin{aligned}
y_k(n)=\mathbf{w}_k^H\left(\mathbf{A}_k(\boldsymbol{\theta})\mathbf{s}(n)+\mathbf{v}_k(n)\right)
\label{subarray_signal}
\end{aligned}
\end{equation}
where $\mathbf{w}_k=[w_{k,1},\cdots,w_{k,M_s}]^T$. $\mathbf{A}_k(\boldsymbol{\theta})=\mathbf{J}_k\mathbf{A}(\boldsymbol{\theta})\in \mathbb{C}^{M_s\times Q}$ is the array steering matrix of $k$th subarray and $\mathbf{J}_k$ is a $M_s\times M$ selection matrix which only contains 0 and 1. The noise vector of the $k$th subarray is also given as $\mathbf{v}_k(n)=\mathbf{J}_k\mathbf{v}(n)$.

As the signal and noise are assumed uncorrelated, the covariance matrix of received signal (\ref{received_signal}) is defined as
\begin{equation}
\begin{aligned}
\mathbf{C}&=\textrm{E}\left[\mathbf{y}(n)\mathbf{y}^H(n)\right]\\
&=\mathbf{W}^H\left(\mathbf{A}\mathbf{C}_s\mathbf{A}^H+\sigma_v^2\mathbf{I}_M\right)\mathbf{W}\\
\label{covariance_matrix}
\end{aligned}
\end{equation}
where $\mathbf{C}_s=\textrm{E}\left[\mathbf{s}(n)\mathbf{s}^H(n)\right]$. However, the covariance matrix $\mathbf{C}$ is usually unavailable in practice, then the sample covariance matrix $\tilde{\mathbf{C}}$ can be employed as an approximation
\begin{equation}
\begin{aligned}
\tilde{\mathbf{C}}&=\frac{1}{N}\sum_{n=1}^N\mathbf{y}(n)\mathbf{y}^H(n)\\
&=\mathbf{C}+\boldsymbol{\varepsilon}
\label{sample_cov_matrix}
\end{aligned}
\end{equation}
where $\boldsymbol{\varepsilon}$ denotes the approximation error.

\begin{figure}[ht]
  \centering
  \includegraphics[width=0.48\textwidth]{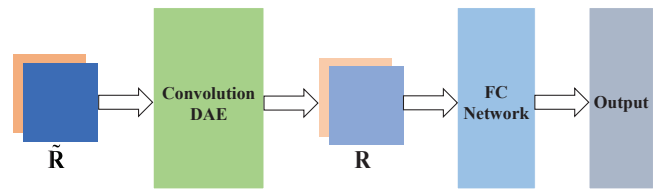}\\
  \caption{Proposed CDAE-DNN.\label{DL_DOA}}
\end{figure}
\section{Deep Learning Based DOA Estimation Method for HAD-OSA}
In this section, we propose a DNN-based DOA estimator which contains convolution DAE and FC network to improve the accuracy of DOA estimation for hybrid massive array with OSA architecture.
\subsection{Data Reprocessing}
To ensure the stability of the input data and improve the accuracy of the neural network model, we choose the sampled covariance matrix $\tilde{\mathbf{C}}$ as the input feature, which is an alternative to the unavailable covariance matrix $\mathbf{C}$. However, the input of neural networks must be real numbers, so we extract both the real part and the imaginary part of $\tilde{\mathbf{C}}$ and construct a $K\times K\times 2$ tensor $\tilde{\mathbf{R}}$, i.e., $\tilde{\mathbf{R}}_{:,:,1}=\textrm{Re}\{\tilde{\mathbf{C}}\}$ and $\tilde{\mathbf{R}}_{:,:,2}=\textrm{Im}\{\tilde{\mathbf{C}}\}$.

Then the label vector $\mathbf{z}=[z_1,z_2,\cdots,z_L]^T$ of input data is defined as follows. Firstly, we assume the angular region containing all the emitters is $[-\theta_0,\theta_0]$, and the label interval is $\Delta\theta$ which is determined by resolution requirement. Therefore, the length of $\mathbf{z}$ is given as $L=\frac{2\theta_0}{\Delta\theta}+1$. And $\mathbf{z}$ is a binary vector containing label 1 at the positions corresponding to the $Q$ training angles and label 0 at the rest positions. So the training dataset can be finally expressed by $\{(\tilde{\mathbf{R}}^{(1)},\mathbf{z}^{(1)}),(\tilde{\mathbf{R}}^{(2)},\mathbf{z}^{(2)}),\cdots,(\tilde{\mathbf{R}}^{(T)},\mathbf{z}^{(T)})\}$.

\subsection{Convolution DAE}
The traditional autoencoder (AE) is a kind of neural network consisting of three parts: encoder, code, and decoder. The input data is first compressed to a lower dimension form, i.e. code, by the encoder, and then the decoder recovers the code to the initial form of the input data. Encoder and decoder have symmetric neural network architectures to perform the opposite operations, so the traditional autoencoders can be summarized as a two-step process
\begin{equation}
\mathbf{r}=f(\tilde{\mathbf{R}})~~ \tilde{\mathbf{R}}=g(\mathbf{r}),
\end{equation}
where $f(\cdot)$ and $g(\cdot)$ denote encode and decode operations respectively, $\mathbf{r}$ represents the code.

Since the input data contains noise $\varepsilon$, the autoencoder can output the noiseless data rather than a simple copy of the input. That is, the two-step process of the autoencoder is transformed to: $\mathbf{r}=f(\tilde{\mathbf{R}})$ and $\mathbf{R}=g(\mathbf{r})$. Otherwise, because the input data is a $K\times K\times 2$ tensor, we consider using the convolution network to implement the function of both the encoder and the decoder. Next, we are going to introduce the complete procedure of the proposed convolutional DAE.

Firstly, assuming the encoder is constructed by a $H$-layers convolution network, the encode function can be modified as
\begin{equation}
\mathbf{r}=f\left(\tilde{\mathbf{R}}\right)=f_H\left(f_{H-1}\left(\cdots f_1\left(\tilde{\mathbf{R}}\right)\right)\right),\label{encoder}
\end{equation}
and each layer contains a convolution layer, a batch normalization (BN) layer and an activation layer. For the $H$ convolution layers, each has $G_h$ filters $h\in\{1,2,\cdots,H\}$. Since the input data is 2-channel, the size of the first convolution layer is $\kappa_1\times\kappa_1\times2\times G_1$. And the sizes of the other $H-1$ convolution layers are given by $\kappa_h\times\kappa_h\times G_h$. Therefore, the output of the $h$th convolution layer can be denoted by $\mathbf{F}_h\in\mathbb{R}^{D_h\times D_h\times G_h}$ and its $u$th channel, $u\in\{1,2,\cdots,G_h\}$, is given by
\begin{equation}
\begin{aligned}
\mathbf{F}_{h,u}=c\left(\mathbf{K}_{h,u},\mathbf{r}_{h-1},\delta_h\right)+\mathbf{b}_{h,u},
\end{aligned}
\end{equation}
where $c(\cdot)$ denotes convolution operation, $\mathbf{K}_{h,u}$ represents the $u$th filter in the $h$th convolution layer. $\mathbf{r}_{h-1}$ is the output of the corresponding layer in the encoder, $\mathbf{r}_0=\tilde{\mathbf{R}}$. $\delta_h$ denotes the stride. $\mathbf{b}_{h,u}$ is the bias matrix of the $u$th filter. The activation function adopted here is RELU, so that the layer output of the encoder can be obtained as
\begin{equation}
\mathbf{r}_{h,u}=\textrm{RELU}\left(\mathbf{F}_{h,u}\right)=\max\left(\textbf{0},\mathbf{F}_{h,u}\right),
\end{equation}
and $\mathbf{r}=\{\mathbf{r}_{H,u}\}_{u=1}^{G_H}$.

Contrary to the encoder, the decoder is required to restore the extracted feature to the form of the original input, which is an upsampling process, also called deconvolution in. Similar to (\ref{encoder}), the decode function is expressed by
\begin{equation}
\hat{\mathbf{R}}=g\left(\mathbf{r}\right)=g_H\left(g_{H-1}\left(\cdots g_1\left(\mathbf{r}\right)\right)\right),
\end{equation}
since the structure of the decoder is symmetric with the encoder, each layer of decoder also contains convolution, BN and activation layers. And the layers' sizes are the same as that of encoder, i.e., $\textbf{\textrm{size}}(g_h)=\textbf{\textrm{size}}(f_{H-h+1})$. It is obvious that in practical application the DAE cannot completely remove the noise $\varepsilon$, so the output of the decoder here is $\hat{\mathbf{R}}$ rather than $\mathbf{R}$.

In the DAE training period, our goal is to find the optimal network parameters based on the training dataset. Thus, we choose MSE as the loss function, and it is defined as
\begin{equation}
\mathcal{L}\left(\Theta\right)=\frac{1}{T}\sum_{i=1}^{T}\left\|\hat{\mathbf{R}}^{(i)}-\mathbf{R}^{(i)}\right\|^2_F,
\end{equation}
where $\Theta$ contains all the weights and biases in the DAE network.

\subsection{Proposed CDAE-DNN}
As shown in Fig.\ref{DL_DOA}, the extracted feature tensor $\tilde{\mathbf{R}}$ is first inputted to the CDAE for eliminating the estimation error $\boldsymbol{\varepsilon}$. Then the output $\hat{\mathbf{R}}$ is inputted to a $(H_{FC}+2)$-layers fully-connected (FC) network. The first layer is a flatten layer, which is used for transforming $\hat{\mathbf{R}}$ into a $2K^2\times1$ vector. And it is followed by $H_{FC}$ dense layers, each containing $G_{h_{FC}}$ neurons, $h_{FC}\in\{1,2,\cdots,H_{FC}\}$. We also choose RELU as the activation function for them, and to achieve regularization in the learning process, the dropout ratio is set as $20\%$. Therefore, the output of $h_{FC}$th dense layer is given as
\begin{equation}
\mathbf{r}_{h_{FC}}=\textrm{RELU}\left(\mathbf{w}_{h_{FC}}\mathbf{r}_{h_{FC}-1}+\mathbf{b}_{h_{FC}}\right),
\end{equation}
where $\mathbf{w}_{h_{FC}}$ and $\mathbf{b}_{h_{FC}}$ denote weight vector and bias vector respectively. When $h_{FC}=1$, $\mathbf{r}_0=\textrm{vec}(\hat{\mathbf{R}})$.

The last layer of the FC network is the output layer with $L$ neurons, and the form of the final output vector is expressed as
\begin{equation}
\hat{\mathbf{z}}=\left[\hat{z}_1,\hat{z}_2,\cdots,\hat{z}_L\right]^T.
\end{equation}
In order to satisfy $0\leq z_l\leq1,~l\in\{1,2,\cdots,L\}$, the activation function for this layer can use sigmoid, which is defined as
\begin{equation}
f(x)=\frac{1}{1+e^{-x}}.
\end{equation}
Then the $Q$ biggest elements are selected from $\hat{\mathbf{z}}$, and their corresponding angles are the estimation results.

Since this is a multi-label problem and we want the final output vectors in the form of probability distributions, we decide to use the binary cross-entropy (BCE) as loss function, which is given by
\begin{equation}
\begin{aligned}
\mathcal{L}_{FC}\left(\Theta_{FC}\right)=-\frac{1}{L}\sum_{l=1}^L &\bigg[z^{(i)}_l\log\left(\hat{z}^{(i)}_l\right)+\\
&\left(1-z^{(i)}_l\right)\log\left(1-\hat{z}^{(i)}_l\right)\bigg],
\end{aligned}
\end{equation}
where $i\in\{1,2,\cdots,T\}$, then the optimal weights and biases of the FC network can be obtained by minimizing it.

\begin{figure}[t]
  \centering
  \includegraphics[width=0.44\textwidth]{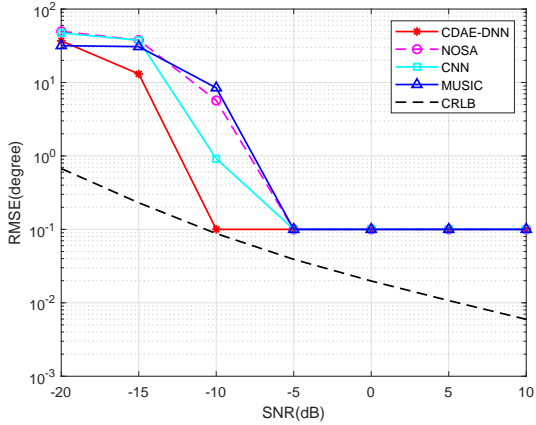}\\
  \caption{RMSE versus SNR.\label{RMSE_SNR}}
\end{figure}
\section{Simulation Results}
In this section, the simulation results are provided for evaluating the performance of the proposed DNN based DOA estimator for the HAD-OSA architecture, and all the simulations related to deep learning are done based on TensorFlow. Firstly, we assume the massive ULA has $M=128$ elements, and the distance between two adjacent elements is $\lambda/2$. Since the OSA architecture is employed in this work, the number of elements in each subarray is $M_s=16$ and the overlapped number is $\Delta M_s=8$. Then the number of RF chains is obtained as $K=15$.  We also suppose the signal source is within the angular range $[-90^{\circ},90^{\circ}]$, and the angular interval is set as $\Delta\theta=1^{\circ}$. In the training period, the CDAE contains one input layer, three convolution layers and one output layer. The FC network contains one flatten layer, three dense layers and one output layer. The three dense layers have 2048, 4096, 2048 neurons respectively, and the output layer also has 181 neurons. The training dataset contains approximately 60000 samples, the batch size and the number of epochs are set as 1000 and 30. Finally, we choose SGD as the optimizer and learning rate is set as 0.1.

Fig.\ref{RMSE_SNR} displays how the DOA estimation accuracy varies with the improvement of SNR. In this simulation, the direction of signal source is set as $\theta=10.1^{\circ}$, the number of snapshots is $N=100$, the range of SNR is -20dB to 10dB, and all the simulation results are averaged over 1000 Monte-Carlo experiments. Except for the proposed CDAE-DNN in this work, we also take three existing methods into consideration as benchmarks. The first is NOSA \cite{shu2018low}, since there is no overlapped region between two adjacent subarrays of it, we let its subarray elements is the same as OSA and the number of RF chains is $K_{\textrm{NOSA}}=8$. The second is a CNN estimator proposed in \cite{papageorgiou2021deep}, which contains four convolution layers and four FC layers. The last is MUSIC \cite{tuncer2009classical}, which is the most popular subspace method for the DOA estimation. Since predicting DOA by using DNN is essentially a multi-classification problem, and the implementation principle of the MUSIC algorithm is also based on the grid search, then there is a lower bound on the estimation accuracy of these methods when the angle to be estimated is off-grid, as shown in Fig.\ref{RMSE_SNR}. This lower bound is dependent on the grid size, which is set as 1 in this simulation, and hence the best estimation RMSE is 0.1. As also can be seen in Fig.\ref{RMSE_SNR}, the proposed CDAE-DNN achieves a significant improvement in estimation accuracy, especially in the low SNR region. Compared with other deep learning-based methods, our method has a greater advancement over the traditional methods. And the comparison with NOSA also proves that OSA can get higher estimation accuracy.

\begin{figure}[t]
  \centering
  \includegraphics[width=0.44\textwidth]{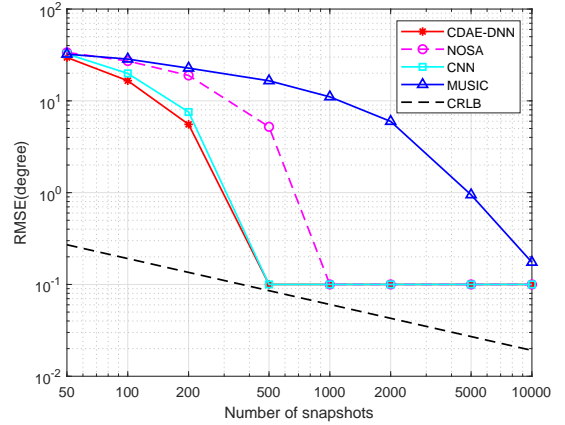}\\
  \caption{RMSE versus the number of snapshot.\label{RMSE_N}}
\end{figure}

Fig.\ref{RMSE_N} shows the relationship between RMSE and the number of snapshots in the environment with SNR=-13dB. The error decreases as N increases and eventually reaches the accuracy lower-bound of 0.1. As can be seen in this figure, our proposed method has great performance advantages under low number of snapshots, especially $N\geq 500$, so it can save a lot of resource overhead compared with the traditional MUSIC algorithm. OSA has also been shown to significantly improve the accuracy of HAD architectures.

\begin{figure}[t]
  \centering
  \includegraphics[width=0.44\textwidth]{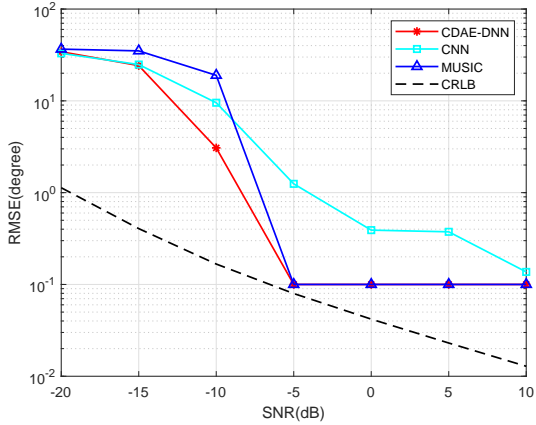}\\
  \caption{RMSE versus SNR, where $Q=2$.\label{rmse_snr_2target}}
\end{figure}

The simulation results in multiple signal source scenarios are given in Fig.\ref{rmse_snr_2target}, where $Q=2$, $\theta_1=10.1^{\circ}$ and $\theta_2=20.1^{\circ}$. The overall trend of the curves is similar to that in the single-target case. The proposed CDAE-DNN has a significant advantage over the MUSIC algorithm, and also has better performance than CNN \cite{papageorgiou2021deep} in medium-high SNR region.
\section{Conclusion}
In this letter, a DL-based DOA estimation method called CDAE-DNN was proposed for the hybrid massive MIMO array with OSA architecture. This estimator was composed of a CDAE and a FC network, the sample covariance matrix was chosen as the input data and its approximation error can be removed by CDAE. The FC network was trained to predict the label of the input signal. Simulation results validated the performance of the proposed  CDAE-DNN, which has significant advantages over traditional MUSIC algorithms and other DL-based methods with low SNR and low snapshots. OSA was also proved to be a reliable option if committed to improving the accuracy of DOA estimation for hybrid array. Finally, the CRLB for the proposed architecture was derived.

\section*{Appendix:~Derivation of CRLB for HAD-OSA}
Referring to the derivation in \cite{friedlander1995direction} and \cite{tuncer2009classical}, the Fisher information matrix (FIM) related to $\boldsymbol{\theta}$ is given as
\begin{equation}
\mathbf{F}=\left[
             \begin{array}{ccc}
               F_{11} & \cdots & F_{1Q} \\
               \vdots & \ddots & \vdots \\
               F_{Q1} & \cdots & F_{QQ} \\
             \end{array}
           \right]
\end{equation}
and its element $F_{\theta_p\theta_q}$ can be expressed as
\begin{equation}
\begin{aligned}
F_{\theta_p\theta_q}=\textrm{tr}\left(\mathbf{C}^{-1}\frac{\partial\mathbf{C}}{\partial\theta_p}\mathbf{C}^{-1}\frac{\partial\mathbf{C}}{\partial\theta_q}\right)
\end{aligned}
\end{equation}
where $1\leq p,q\leq Q$, $\mathbf{C}$ is the covariance matrix of received signal and
\begin{equation}
\begin{aligned}
\frac{\partial\mathbf{C}}{\partial\theta_q}&=\frac{\partial\tilde{\mathbf{A}}}{\partial\theta_q}\mathbf{C}_s\tilde{\mathbf{A}}^H+\tilde{\mathbf{A}}\mathbf{C}_s{\frac{\partial\tilde{\mathbf{A}}}{\partial\theta_q}}^H\\
&=\mathbf{W}^H\mathbf{D}_q\mathbf{C}_s\tilde{\mathbf{A}}^H+\tilde{\mathbf{A}}\mathbf{C}_s\mathbf{D}_q^H\mathbf{W}\\
&=\mathbf{W}^H\mathbf{D}\mathbf{e}_q\mathbf{e}_q^T\mathbf{C}_s\tilde{\mathbf{A}}^H+\tilde{\mathbf{A}}\mathbf{C}_s\mathbf{e}_q\mathbf{e}_q^T\mathbf{D}^H\mathbf{W}
\end{aligned}
\end{equation}
where $\tilde{\mathbf{A}}=\mathbf{W}^H\mathbf{A}$, $\mathbf{e}_q$ denotes the $q$th column of identity matrix $\mathbf{I}_Q$ and
\begin{subequations}
\begin{align}
&\mathbf{D}=\sum_{q=1}^Q\mathbf{D}_q,\\
&\mathbf{D}_q=\left[\textbf{0}_{M\times (q-1)}~~\mathbf{d}_q\mathbf{a}(\theta_q)~~\textbf{0}_{M\times (Q-q)}\right],\\
&\mathbf{d}_q=\textrm{diag}\left\{0,j\frac{2\pi}{\lambda}d\cos\theta_q,\cdots,j\frac{2\pi}{\lambda}(M-1)d\cos\theta_q\right\},
\end{align}
\end{subequations}
according to the equation $\textrm{tr}(\mathbf{A}^H)=\textrm{tr}(\mathbf{A})^{\ast}$ we can get
\begin{equation}
\begin{aligned}
F_{\theta_p\theta_q}=&2\textrm{Re}\bigg\{\textrm{tr}\left(\mathbf{W}^H\mathbf{D}_p\mathbf{C}_s\tilde{\mathbf{A}}^H\mathbf{C}^{-1}\mathbf{W}^H\mathbf{D}_q\mathbf{C}_s\tilde{\mathbf{A}}^H\mathbf{C}^{-1}\right)\\
&+\textrm{tr}\left(\mathbf{W}^H\mathbf{D}_p\mathbf{C}_s\tilde{\mathbf{A}}^H\mathbf{C}^{-1}\tilde{\mathbf{A}}\mathbf{C}_s\mathbf{D}_q^H\mathbf{W}\mathbf{C}^{-1}\right)\bigg\},
\end{aligned}
\end{equation}
therefore, by combining all the elements in $\mathbf{F}$ can obtain
\begin{equation}
\begin{aligned}
\mathbf{F}=&2\textrm{Re}\bigg\{\left(\mathbf{C}_s\tilde{\mathbf{A}}^H\mathbf{C}^{-1}\mathbf{W}^H\mathbf{D}\right)\odot\left(\mathbf{C}_s\tilde{\mathbf{A}}^H\mathbf{C}^{-1}\mathbf{W}^H\mathbf{D}\right)^T\\
&+\left(\mathbf{C}_s\tilde{\mathbf{A}}^H\mathbf{C}^{-1}\tilde{\mathbf{A}}\mathbf{C}_s\right)\odot\left(\mathbf{D}^H\mathbf{W}\mathbf{C}^{-1}\mathbf{W}^H\mathbf{D}\right)^T\bigg\},
\end{aligned}
\end{equation}
and by collecting the signals at all the $N$ snapshots, the CRLB is given as
\begin{equation}
\textrm{CRLB}=\frac{1}{N}\mathbf{F}^{-1}.
\end{equation}

\ifCLASSOPTIONcaptionsoff
  \newpage
\fi

\bibliographystyle{IEEEtran}
\bibliography{DOA_OSA_ref}
\end{document}